\documentclass[12pt]{article}
\usepackage{epsfig}
\usepackage{float}
\bibliography{plain}

\title{\bf The $\eta$-pseudo-hermitic generator in the deformed Woods Saxon’s potential
 }
\author{M. Hafezghoran, Z. Bakhshi\thanks{Corresponding author (E-mail: z.bakhshi@shahed.ac.ir)}\\
{\small Department of Physics, Faculty of Basic Sciences, Shahed University, Tehran, Iran.} \\ }\pagebreak
\begin{document}
\maketitle
\vspace{-10mm}
\begin{abstract}
In this paper, we present a general method to solve non-hermetic potentials with PT symmetry using the definition of two $\eta$-pseudo-hermetic and first-order operators. This generator applies  to the Dirac equation which consists of two spinor wave functions and non-hermetic potentials, with position that mass is considered a constant value and also Hamiltonian hierarchy method and the shape invariance property are used to perform calculations. Furthermore, we show the correlation between the potential parameters with transmission probabilities that $\eta$-pseudo-hermetic using the change of focal points on Hamiltonian can be formalized based on Schr\"{o}dinger-like  equation.  By using this method for some solvable potentials such as deformed Woods Saxon's potential, it can be shown that these real potentials can be decomposed into complex potentials consisting of eigenvalues of a class of $\eta$-pseudo-hermitic.
\end{abstract}
\pagebreak \vspace{7cm} \pagebreak \vspace{7cm}
\section{Introduction}
Relativistic quantum mechanics has always been considered as an important topic about the effects of particles by potential fields. In order to study fermion particles in some physics branches such as modern physics and particle physics, we need to solve Dirac equation. Non-hermetic models play an important role in physics systems such as nuclear physics, quantum field theory, and more. In recent years, various methods have been proposed to describe the solution of relativistic and non-relativistic equations with the help of non-hermeneutic Hamiltonians, which consist of two real and imaginary parts. Therefore, in this paper, by presenting $\eta$-pseudo-hermetic operators, we have solved the Dirac equation in a situation where the constant mass is considered, which can be interesting. Since $\eta$ does not depend on hermetic, linear or reversible requisite, we can rewrite Hamilton as pseudo-hermitians. In fact since the Dirac equation is divided into two first-order differential spins of the upper and lower spins, by separating the components, the second component can be rewritten as the first component. So we have presented that the Dirac equation will be solved by formulating a Schr\"{o}dinger-like equation that consists of defining a pseudo-hermetic operator to create a class of PDEM Hamiltonians while a constant mass is considered. On the other hand in past decades, variant solutions were presented for Dirac equation [1-3]. For example we can name shape-invariant method in super symmetric quantum mechanics. Supper symmetric quantum mechanics, for solvable  potentials, makes it for eigenvalues and eigenfunctions to be obtained using symmetric operator formulation. In fact in field theory and by applying $d=1$ limitation (one dimensional) we can use supper symmetric in quantum mechanics [4]. Gendenshtein [5] was a pioneer in offering shape-invariant potentials in quantum mechanics method so that, in recent years, many solvable problems related to this field has been solved using shape-invariant potentials. Beside that, super symmetric in quantum mechanics is defined based on upon the factorization method in shape-invariant limitation. If one quantum mechanics consists of super symmetry contents thus we factorize the Hamiltonian equation based on first-order differential operators in shape-invariance equations.\\
In this approach, Hamiltonian is once decomposed in successive multiplication for raising and lowering operators and then in successive multiplication of lowering and raising so that the corresponding quantum status of successive levels, are their eigenstates. These Hamiltonians are each other's partner and super symmetric. Also in the field theory discussion, there are different solvable potentials that we want to use one of them is called the Woods-Saxon potential to solve the problem in the Dirac equation. The spherical Woods-Saxon potential as a well-known model in different physics brunches such as nuclear physics, has been the center of attention. For example this potential in microscopic fields has had an important role as a barrier and well in central and symmetrical parts between a Neutron and heavy Nucleus [6,7]. Woods-Saxon potential used as the major part of Nuclear shell model successfully deduced the nuclear energy level also it is used as the main part of Neutron interaction with heavy Nucleus [8]. Using the axially deformed Woods-Saxon potential with spin-orbit interaction potential we are able to create the structure of single-particle shell model [9]. The Woods-Saxon potential is used as a part of optical model in elastic scattering of some ions with heavy targets in low range of energies. Overall, Woods-Saxon potentials and its different shape models was successful to describe the metallic clusters [10]. In recent decades, Dirac relativistic equation was solved using to component spinors for Woods-Saxon potential in special circumstances . Just like Woods-Saxon potential and even in its modified models, none-hermitic models were of great importance for the scientists in different brunches of physics such as nuclear physics. In order to define the solution of relativistic and none-relativistic equation, non-hermitic Hamiltonian consisting of real or imagining spectra can be used. Therefore, the solution of diminished Dirac equation with constant mass can be used to solve constant mass equations with the same approach [11].\\
Non-Hermitic Hamiltonians always contain the necessity of hermetic parts, thus in order to study the real value of energy spectrum using the $PT$ symmetry method, hermetic Hamiltonians are required for none hermetic Hamiltonian based on $PT$ symmetry we will have $\hat{O}\hat{H}\hat{O^{-1}}=\hat{O^{-1}}\hat{H}\hat{O}=H$ in which the $\hat{O}=PT$ operator is a combined operator of Parity and time inversion transformations. If a potential is $\nu(x)=\nu^{*}(x)$ sub-conversion $i \to -i$ and $x \to -x$, we consider it a potential with $PT$ symmetry [12].\\
At the moment more generalized method is offered to prove if an energy spectrum is true, and it indicates $PT$ symmetry Hamiltonian consist of sub-systems named Pseudo-Hermetic. This means Pseudo-Hamilton is the original Hamilton if follow the conversations written below :\\
 $\hat{\eta}\hat{H}\hat{\eta^{\dagger}}=\hat{\eta^{\dagger}}\hat{H}\hat{\eta}=H$  \\ 
in which $(\dagger)$ is an accessory marker and $\eta$ is a hermetic inverse linear operator and $\eta=O^{\dagger}O , O:\mathcal{H} \to \mathcal{H}$ ($\mathcal{H}$ is Hilbert Space) [13].\\
So it can be planed that $PT$ symmetric or hermetic symmetric, is not of enough necessity to verify the existence of a quantum Hamiltonia. In fact, three separate subjects i.e. factorization method, super symmetry in quantum mechanics and shape-invariant method is converged at one point. Deformed Woods-Saxon $(DWS)$ potential is a short range potential, generally used in nuclear physics, particle physics, atomic condensed matter and chemical physics [14]. The factorization of the spin-orbit part of the potential is obtained in the region corresponding to large deformations (second minima) depending only on the nuclear surface area. The spin-orbit interaction of a particle in a non-central self consistent field of the deformed Woods-Saxonl potential model is investigated for light nuclei and the scheme of single-particle states has been found for mass numbers $A_{0}$ = 10 and 25 [15]. Woods-Saxon potential is evaluated in Schr\"{o}dinger and Klein Gordon equation. In this paper, we tended to evaluated $PT$ and non-$PT$-symmetric relativistic equation and non-Hermitian modified Woods–Saxon potential. Also we want to investigated relateable answers to Dirac equation for Deformed Woods-Saxon potential considering real and imaginary energy levels [16]. Woods-Saxon potential can be studied using the studies of single-particle structure for Plutonium odd isotopic based on parameterization for spin-orbit interactions . Also, using the Wood-Saxon form in optical potentials, we investigate the possibility of measurement of differential cross-section in some energies in the elastic scattering.\\
In recent years , spin and pseudo-spin symmetry concepts were introduced in nuclear theory [17]. In order to indicate features of deformed nuclei  and super deformation and to establish a shell model coupling scheme [18]. In fact, within relativistic mean field-theory Ginocchio found that a Dirac Hamiltonian has a spinor symmetry along with a $U(3)$ symmetry and a scalar potential and a repulsive vector potential have the same magnitudes which means $S(r) = V(r)$. Furthermore when $S(r) = -V(r)$ contains a super symmetry like pseudo-$U(3)$ symmetry for instance, harmonic oscillator [19], the Woods-Saxon potential [20], Hartmann potentials [21] and the Morse potential [22] can be named that they were investigated.\\
So in this paper Dirac equation is investigated using super symmetric quantum mechanics method, shape-invariant theory, the concept of defining pseudo-hermitic operators on the issue of problem Hamiltonian and in correlation with the bound state energy eigenvalues and the corresponding spinor wave functions are calculated.

\section{ The one particle Dirac equation in one dimension }
\setcounter{equation}{0}
In one dimension Dirac equation answers are divided in two parts. So that negative and positive energy are stable without considering complications of spin. Starting with the relativistic free particle Dirac equation $(\hbar=c=1)$ :      
\begin{equation}
(i\gamma^{3}\frac{\partial}{\partial x^{3}}-M)\Psi(x)=0\end{equation}
In presence of on external potential $V(x)$ and taking the gamma matrices can considering $\gamma_{x}$ and $\gamma_{0}$ as $i\sigma_{x}$ and $\sigma_{z}$ the Pauli matrices respectively $(\mu=1, 2, 3)$ , one dimension Dirac equation is written as follows:   
\begin{equation}
[\sigma_{x}\frac{d}{dx}-(E-V(x))\sigma_{z}-M]\Psi(x)=0\end{equation}
Four-spinour $\Psi$ are broken down to two spinors $u_{1}$ and $u_{2}$, we have:
\begin {equation}
\frac{d}{dx}\left(\begin{array}{c}
       u_{1} \\ u_{2}
      \end{array}\right)-(E-V(x))\left(\begin{array}{c}
                   -u_{2} \\ u_{1}
                 \end{array}\right)+M\left(\begin{array}{cc}
                    0 & 1 \\ 1 & 0 \end{array}\right)\left(\begin{array}{c}
                   u_{2} \\ u_{1}
                 \end{array}\right)=0\end{equation}
And this problem is solvable using two differential equations: 
\begin{equation}
u^{\prime}_{1}(x)+(E-V(x))u_{2}(x)+Mu_{2}(x)=0\end{equation}
\begin{equation}
u^{\prime}_{2}(x)-(E-V(x))u_{1}(x)+Mu_{1}(x)=0\end{equation}
We use a method similar to Fl\"{u}gge [23] method and two combination are defined:
\begin{equation}
\Phi(x)=u_{1}(x)+iu_{2}(x) \qquad\qquad\qquad,\qquad\qquad\qquad \chi(x)=u_{1}(x)-iu_{2}(x) \end{equation}
By replacing obtained combinations in Eq.(2.4) and Eq.(2.5) satisfy followed equations:
\begin{equation}
\Phi^{\prime}(x)=-iM\chi(x)+i(E-V(x))\Phi(x) \end{equation}
\begin{equation}
\chi^{\prime}(x)=iM\Phi(x)-i(E-V(x))\chi(x) \end{equation}
By eliminating bottom spinor factor in $\chi(x)$ two above equations and combining this equations, we can write a Schr\"{o}dinger-like equation with constant mass for up spinor factor $\Phi(x)$:
\begin{equation}
-\frac{d^{2}\Phi(x)}{dx^{2}}+[2\varepsilon\nu(x)-\nu^{2}(x)-i\frac{d\nu(x)}{dx}-\varepsilon^{2}+M^{2}]\Phi(x)=0 \end{equation}

\section{ Super Symmetric quantum mechanics equations }
\setcounter{equation}{0}
In $N=2$ super symmetric quantum mechanics $(SQM)$, it is necessary to define two nilpotent operators namely $Q$ and $Q^{+}$, satisfying the algebra:
\begin{eqnarray}
\left\{Q, Q^{+}\right\}=H_{s}   \nonumber \\ \left\{Q, Q\right\}=\left\{Q^{+}, Q^{+}\right\}=0 \end{eqnarray}
where $H_{s}$ is the super symmetric Hamiltonian. These operators can be realized as:
\begin{equation}
Q=\left(\begin{array}{cc}
             0 & 0 \\ \eta^{-} & 0 \end{array}\right) \qquad\qquad\qquad,\qquad\qquad\qquad 
Q^{+}=\left(\begin{array}{cc}
             0 & \eta^{+} \\ 0 & 0 \end{array}\right) \end{equation}
where $\eta^{+}$ and $\eta^{-}$ are bosonic operators. The Hamiltonian, $H_{s}$ in terms of these operators is given by:
\begin{equation}
H_{s}=\left(\begin{array}{cc}
            \eta^{+}\eta^{-} & 0 \\ 0 & \eta^{-}\eta^{+} \end{array}\right)=\left(\begin{array}{cc}
             H^{-} & 0 \\ 0 & H^{+} \end{array}\right) \end{equation}
In the other hand $H^{\pm}$ are called super symmetric partner Hamiltonians and share the same spectra, apart from the non degenerate ground state, (see [9] for a review):
\begin{equation}
E^{(+)}_{n}=E^{(-)}_{n+1} \end{equation}

\section{ $\eta$-pseudo-hermitic operators and Schr\"{o}dinger-like equation  }
\setcounter{equation}{0}
Two first-order hermitic and non-hermitic operators whit constant mass $(\mu=\frac{1}{M})$ will be introduced as below [24]:
\begin{equation}
\eta^{+}=-i\mu\frac{d}{dx}+F(x) \end{equation}
\begin{equation}
\eta^{-}=\mu\frac{d}{dx}+iF(x) \end{equation}
The operators $\eta^{\pm}$ are defined in terms of the superpotential $F(x)$, in which the value of $\mu$ is constant and it doesn't depend on another parameter. $Z=\mu^{2}\frac{d}{dx}[i(\varepsilon-\nu(x))]$ and $V(x)=-\mu^{2}(\varepsilon-\nu(x))$ phrases are named and equation $\eta^{\pm}H=H^{\dagger}\eta^{\pm}$ in replaced in equation (2.9).\\
After parsing and simplifying the imaginary part of $\partial_{x}\Psi(x)$ we have :
\begin{equation}
2i\mu Z=-2i\mu^{2}F^{\prime}(x) \end{equation}
So we have the imaginary part of potential as follows: 
\begin{equation}
Z(x)=-\mu F^{\prime}(x) \end{equation}
and then:
\begin{equation}
-i\mu^{2}F^{\prime\prime}(x)=i\mu Z^{\prime}(x) \end{equation}
\begin{equation}
2\mu^{3}(\varepsilon-\nu(x))\nu^{\prime}(x)-Z(x)F(x)=Z(x)F(x) \end{equation}
from the real part of $\Psi(x)$, the real potential phrases in operator $F(x)$ is as follows:
\begin{equation}
V(x)=F^{2}(x)+\alpha_{0} \end{equation}
\begin{equation}
V_{\pm}(x)=F^{2}(x)\pm i\mu F^{\prime}(x)+\alpha_{0} \end{equation}
and $\alpha_{0}\in R$ is integration constant, by definition two partner potentials are called shape invariant if they have the same functional form, differing only by change of parameters, including an additive constant. In this case the partner potentials satisfy:
\begin{equation}
V_{+}(x, a_{1}) = V_{-}(x, a_{2})+R(a_{2}) \end{equation}
where $a_{1}$ and $a_{2}$ denote a set of parameters, with $a_{2}$ being a function of $a_{1}$:
\begin{equation}
a_{2} = f(a_{1}) \end{equation}
and $R(a_{2})$ is independent of $x$.\\
For the non-spontaneously broken supersymmetry this lowest level is of zero energy, $E^{(1)}_{0}=0$. We have :
\begin{equation}
H^{\pm}=-\mu^{2}\frac{d^{2}}{dx^{2}}+V_{\pm}(x)=\eta^{\mp}\eta^{\pm} \end{equation}
Using the algebra, for $H_{1}$ special Hamiltonians which is a factor of  bosonic operators, a hierarchy of Hamiltonian can be written. In this case our spontaneous symmetric is broken as follows :
\begin{equation}
H_{1}=-\mu^{2}\frac{d^{2}}{dx^{2}}+V_{1}(x)=\eta^{+}_{1}\eta^{-}_{1}+E^{(1)}_{0} \end{equation}
so that $E^{(1)}_{0}$ is the minimum energy level.\\
Bosonic operators we defined in Eq.(4.1), Eq.(4.2) so that $F(x)$ super potential applies in Riccati equation :
\begin{equation}
V_{1}(x)=-F^{2}_{1}(x)-i\mu F^{\prime}_{1}(x)+E^{(1)}_{0} \end{equation}
for lower energy states, eigenfunction is associated with $F(x)$ super potential:
\begin{equation}
\Phi^{(1)}_{0}(x)=N exp(\frac{-1}{\mu}\int_{0}^{x}F_{1}(\bar{x})d\bar{x}) \end{equation}
also super symmetry partner Hamiltonian is calculated as follows: 
\begin{equation}
H_{2}=\eta^{-}_{1}\eta^{+}_{1}+E^{(1)}_{0}=-\mu^{2}\frac{d^{2}}{dx^{2}}+(-F^{2}_{1}-i\mu F^{\prime}_{1})+E^{(1)}_{0} \end{equation}
therefor $H_{2}$ Hamiltonian can be factorized as couple-terms of bosonic operators and thus for $\eta^{\pm}_{2}$ we will have: 
\begin{equation}
H_{2}=\eta^{+}_{2}\eta^{-}_{2}+E^{(2)}_{0}=-\mu^{2}\frac{d^{2}}{dx^{2}}+(-F^{2}_{2}-i\mu F^{\prime}_{2})+E^{(2)}_{0} \end{equation}
so that $E^{(2)}_{0}$ is the minimum eigenvalue of $H_{2}$ and applies in Riccati equation:
\begin{equation}
V_{2}(x)=-F^{2}_{2}-i\mu F^{\prime}_{2}+E^{(2)}_{0} \end{equation}
Hence, using the eigenvalue and eigenfunction of a set of $n$-members a related continuum of Hamiltonians can be written :
\begin{equation}
H_{n}=\eta^{+}_{n}\eta^{-}_{n}+E^{(n)}_{0} \end{equation}
\begin{equation}
\eta^{+}_{n}=-i\mu\frac{d}{dx}+F_{n}(x) \qquad\qquad , \qquad\qquad \eta^{-}_{n}=\mu\frac{d}{dx}+iF_{n}(x) \end{equation}
\begin{equation}
\Phi^{(1)}_{n}=\eta^{+}_{1}\eta^{+}_{2}\cdots\eta^{+}_{n}\Phi^{(n+1)}_{0} \qquad\qquad , \qquad\qquad E^{(1)}_{n}=E^{(n+1)}_{0} \end{equation}
so that $\Psi^{(1)}_{0}$ is obtained from (4.14).

\section{ Deformed Woods-Saxon  potential  }
\setcounter{equation}{0}
Followed potential is considered as general form of deformed Woods-Saxon  potential:
\begin{equation}
V(x)=\frac{-V_{0}e^{-(\frac{x-X_{0}}{a})}}{1+qe^{-(\frac{x-X_{0}}{a})}}+\frac{C_{0}e^{-2(\frac{x-X_{0}}{a})}}{({1+qe^{-(\frac{x-X_{0}}{a})}})^{2}} \end{equation}
$x$ indicates the distance between the mass center of target nucleus and the projectile nucleus. Other parameters in equation $q$ (as deformed parameter for $q\ge 1$) and $X_{0}=x_{0}A^{1/3}$ is as radius of spherical nucleus or the expansion of potential. $A$ is numerical limitation of mass particle, $x$ is a radial parameter (remember in this article, radial parameter related $r$, values is considered $x$ due to the imitation of the general linear form of the Woods-Saxon potential).\\
$V_{0}$ indicates potential depth and $a$ is diffuseness of the surface and $c$ is parameter value regulator, which is defined by us. Therefore, continuum Hamiltonians can be formalized based on Schr\"{o}dinger-like equation:
\begin{equation}
-\mu^{2}\frac{d^{2}}{dx^{2}}\Phi(x)+V(x)\Phi(x)=E\Phi(x) \end{equation}
and the ground state energy is written as follows:
\begin{equation}
\Phi_{0}(x)=N exp(\frac{-1}{\mu^{2}}\int F_{1}(x)dx) \end{equation}
where $N$ is normalization constant. By replacing (5.3) in (5.2) we will have:
\begin{equation}
V_{1}(x)=-F_{1}^{\prime\prime}(x)-i\mu F_{1}^{\prime}(x)+E_{0}^{(1)} \end{equation}
so that $E_{0}^{(1)}$ is minimum eigenvalues or in either words is the grand state energy. Using algebra equation, super potential is defined as follows: 
\begin{equation}
F_{1}=-\mu(G_{1}+G_{2}\frac{e^{-\alpha(x-X_{0})}}{1+qe^{-\alpha(x-X_{0})}}) \end{equation}
which applies in Riccati equation and by replacing (5.5) in equation (5.4) following values will be obtained:
\begin{eqnarray}
\mu^{2}G_{1}^{2}+\mu^{2}(\frac{2G_{1}G_{2}-i\alpha G_{2}}{q+e^{\alpha(x-X_{0})}})+\mu^{2}\frac{(G_{2}^{2}+i\alpha qG_{2})}{(q+e^{\alpha(x-X_{0})})}=V_{1}(x)-E_{0}^{(1)} \nonumber\\
=\frac{-V_{0}}{q+e^{(x-X_{0})/a}}+\frac{c}{(q+e^{(x-X_{0})/a})^{2}}-E_{0}^{(1)} \end{eqnarray}
By comparing three sections of the each side of Eq.(5.6), we will have:
\begin{eqnarray}
\alpha=1/a \nonumber \\  \mu^{2}G_{1}^{2}=-E_{0}^{(1)} \nonumber \\ \mu^{2}(2G_{1}G_{2}-i\alpha G_{2})=-V_{0} \nonumber \\ \mu^{2}(G_{2}^{2}+i\alpha qG_{2})=c \end{eqnarray}
$\Phi_{0}(x)$ eigenfunction for ground state is obtained as follows:
\begin{equation}
\Phi_{0}(x)=N exp[\int(G_{1}+G_{2}\frac{e^{-\alpha(x-X_{0})}}{1+qe^{-\alpha(x-X_{0})}})dx]=N e^{G_{1}x}(\frac{e^{\alpha(x-X_{0})}}{e^{-\alpha(x-X_{0})}+q})^{\frac{G_{2}}{\alpha q}} \end{equation}
By solving the equation (5.7), we get:
\begin{eqnarray}
G_{1}=\frac{(-V_{0}+c/q)}{2G_{2}\mu^{2}}-\frac{G_{2}}{2q} \nonumber \\ G_{2}=-i\frac{\alpha q}{2}\pm\sqrt{(\frac{\alpha q}{2})^{2}+\frac{c}{\mu^{2}}} \end{eqnarray}
Here we use the equations (5.9) and (5.5) the supersymmetric potential pair can be obtained as follows:
\begin{eqnarray}
V_{+}(x)=\mu^{2}[G_{1}^{2}+\frac{\frac{1}{\mu^{2}}(-V_{0}+\frac{c}{q})-\frac{2G_{2}^{2}}{q}}{q+e^{\alpha(x-X_{0})}}+\frac{G_{2}^{2}}{(q+e^{\alpha(x-X_{0})})^{2}} \nonumber\\
+i\frac{\alpha G_{2}}{q+e^{\alpha(x-X_{0})}}-i\frac{\alpha qG_{2}}{(q+e^{\alpha(x-X_{0})})^{2}}] \end{eqnarray}
\begin{eqnarray}
V_{-}(x)=\mu^{2}[G_{1}^{2}+\frac{\frac{1}{\mu^{2}}(-V_{0}+\frac{c}{q})-\frac{2G_{2}^{2}}{q}}{q+e^{\alpha(x-X_{0})}}+\frac{G_{2}^{2}}{(q+e^{\alpha(x-X_{0})})^{2}} \nonumber\\
-i\frac{\alpha G_{2}}{q+e^{\alpha(x-X_{0})}}+i\frac{\alpha qG_{2}}{(q+e^{\alpha(x-X_{0})})^{2}}] \end{eqnarray}
considering the realation between $V_{+}$ and $V_{-}$ we have:
\begin{equation}
V_{+}(x, G_{2})=V_{-}(x, G_{2}-\alpha q)+\mu^{2}[\frac{-V_{0}+c/q}{2\mu^{2}G_{2}}-\frac{G_{2}}{2q}]^{2}-\mu^{2}[\frac{-V_{0}+c/q}{2\mu^{2}(G_{2}-\alpha q)}-\frac{G_{2}-\alpha q}{2q}]^{2} \end{equation}
also considering the concept of shape-invariance by Gendenshtein [1,2] and by comparison Eq.(5.12)  and Eq.(4.9) we obtained:\\  
 $G_{2}\longrightarrow a_{1}$ \\ $G_{2}-\alpha q\longrightarrow a_{2}$
\begin{equation}
R(a_{2})=\mu^{2}[\frac{-V_{0}+c/q}{2\mu^{2}a_{1}}-\frac{a_{1}}{2q}]^{2}-\mu^{2}[\frac{-V_{0}+c/q}{2\mu^{2}a_{2}}-\frac{a_{2}}{2q}]^{2} \end{equation}
based on shape-invariance concept, that we define as residual $R(a_{2})$ is independent of $x$ value and thus using the Hamiltonian equation we will have:
\begin{equation}
H^{(k)}=-\mu^{2}\frac{d^{2}}{dx^{2}}+V_{-}(x, a_{k})+\Sigma_{a=1}^{k}R(a_{k}) \end{equation} 
so that $H^{(k)}$ is written as continuum Hamiltonian in which $k=1, 2, 3, ...$ , $H^{(1)}\equiv H_{0}$. There are $H^{(k)}$ energy spectrum is calculated as:
\begin{equation}
E_{0}^{(k)}=\Sigma_{a=1}^{k}R(a_{k}) \end{equation}
By using Eq.(5.15) we can obtain level $n$ energy values of Hamiltonian related to the ground state $H_{n} (n=0,1,2,...)$ and also the energy eigenvalue of Hamiltonian are given by:
\begin{equation}
E_{0}^{(-)}=0 \end{equation}
\begin{equation}
E_{n}^{(-)}=\mu^{2}[\frac{-V_{0}+c/q}{2\mu^{2}G_{2}}-\frac{G_{2}}{2q}]^{2}-\mu^{2}[\frac{1}{2\mu^{2}(G_{2}-n\alpha q)}-\frac{G_{2}-n\alpha q}{2q}]^{2} \end{equation}
and this is when deformed Woods-Saxon potential in equation (5.1) for the zero angular momentum are found as: 
\begin{equation}
E_{n}=E_{n}^{(-)}+E_{0}^{(-)}=-\mu^{2}[\frac{1}{2\mu^{2}(G_{2}-n\alpha q)}-\frac{G_{2}-n\alpha q}{2q}]^{2} \end{equation}
Substituting Eq.(5.9) into Eq.(5.18) and considering $c=0$ and $G_{2}=-\alpha q$ the quantity difference between $V(x)$ and $V_{-}(x)$ (as residual values) associated with ground state energy changes $V(x)=V_{-}(x)+E_{0}$, eigenvalue of energy is obtained as follows:
\begin{equation}
E_{n}=-\frac{\mu^{2}}{a^{2}}[(\frac{a^{2}V_{0}}{\mu^{2}q(n+1)})^{2}+(\frac{n+1}{2})^{2}+\frac{2aV_{0}^{2}}{\mu^{2}q^{2}}] \end{equation}
the negative sign $G_{2}$ is associated with $q>0$ values, these result on form confirmation graphs, is energy eigenvalue [25].\\
   Eq.(4-19) is similar to result calculated before applying the factorization method [23], super symmetric approach [24], quasi-linearization method [25] and Nikiforov-Uvarov method [26,27].\\
We have the bound state energy eigenvalues $E_{n,l}(l=0)$ for some values of $n$ with $X=1.25A_{0}^{1/3} /fm , V_{0}=40.5+0.13A_{0} , a=0.65 , \hbar=6.5821\times10^{-22}MeV.s , A_{0}=40$ .
\begin{figure}[H]
\centerline{\includegraphics[scale=0.6]{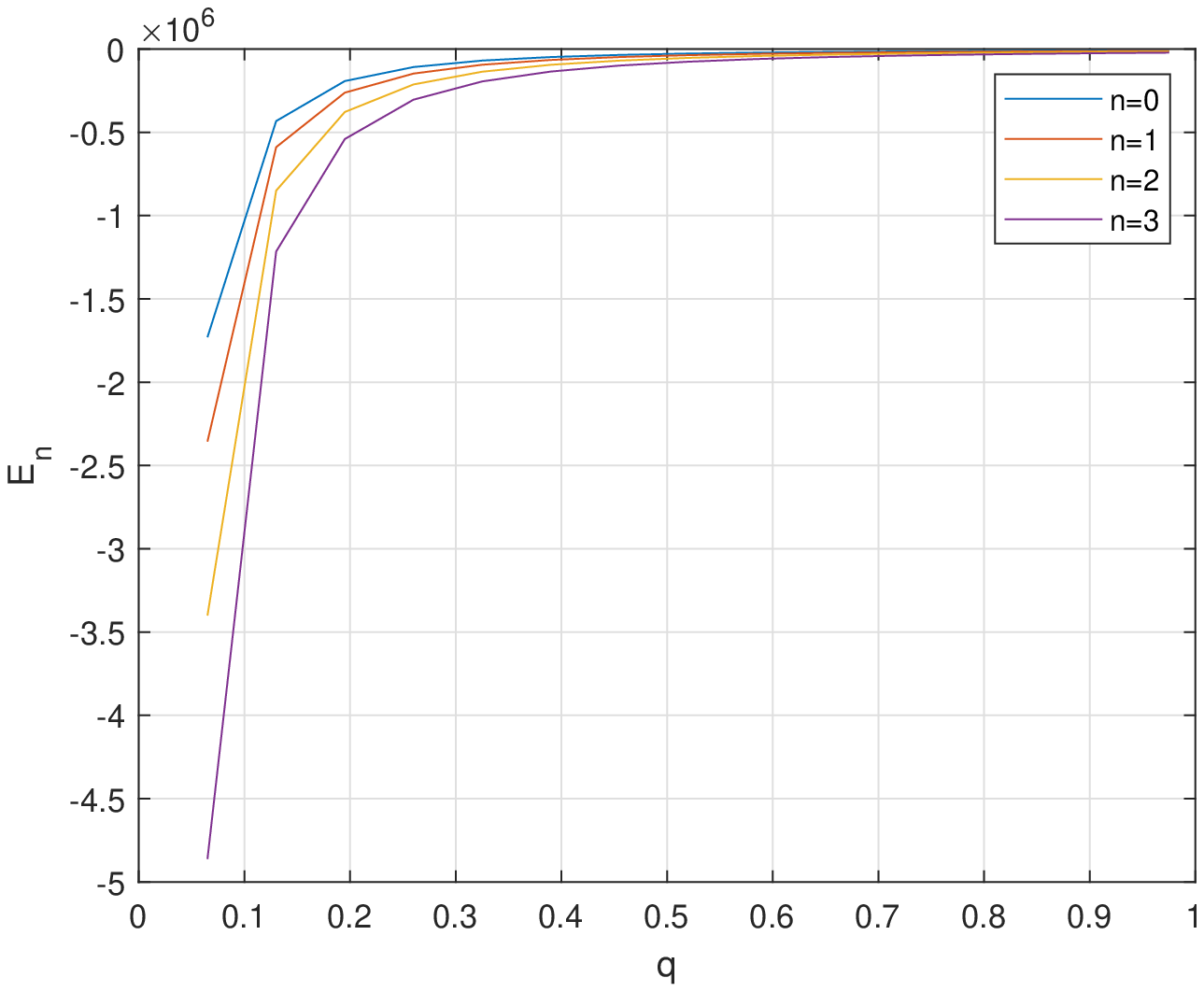}}
\caption{\scriptsize{The variation of the ground state $(n=0)$ and $(n=1,2,3)$ energy level as a function of the deformation parameter. We choose $\mu=1 fm^{-1}, a=0.65 fm, X=1.25A_{0}^{1/3}, V_{0}=40.5+0.13A_{0} /fm, A_{0}=40.$}}
\label{fig1}
\end{figure}
To discuss the variant behavior of energy spectrum with the deformation parameter $q$, ground state is illustrated for $n=0$, first, second and third excited $(n=1,2,3)$ and $q$ variation spectrum is drawn in Figs.1.\\
In Figs.1, when deformation parameter $q$ increase, particle is less attracted or its energy is less negative (tends to continuum states). Nonetheless, when $q<0.1$ the particle is strongly bound since the increasing $q$ shields the $WS$ field.\\
\begin{figure}[H]
\centerline{\includegraphics[scale=0.6]{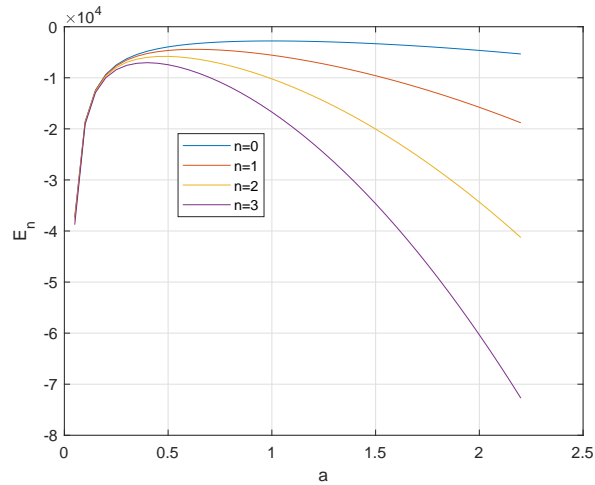}}
\caption{\scriptsize{The variation of the ground state $(n=0)$ and $(n=1,2,3)$ energy level as a function of the deformation parameter. We choose $\mu=1 fm^{-1}, q=1.5 fm, X=1.25A_{0}^{1/3}, V_{0}=40.5+0.13A_{0} /fm, A_{0}=40.$}}
\label{fig1}
\end{figure}
In Figs.2, we also plot the ground $n=0$ and $n=1,2,3$ excited states. We see that for the case $q=1.5 fm$, the energy curve is strongly bound for a wide range of $"a"$ and although the depth of energy curve for marked sections is short but its difference is obvious when on a smaller scale, the energy is investigated.\\
\begin{figure}[H]
\centerline{\includegraphics[scale=0.6]{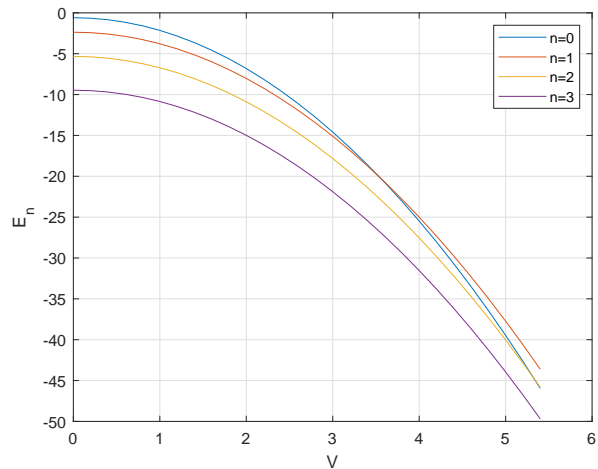}}
\caption{\scriptsize{The variation of the ground state $(n=0)$ and $(n=1,2,3)$ energy level as a function of the potential depth $V_{0}$. We choose $\mu=1 fm^{-1}, q=1.5 fm, a=0.65 fm, X=1.25A_{0}^{1/3}, A_{0}=40.$}}
\label{fig1}
\end{figure}
Finally in Figs.3, we draw the energy $E_{n}$ against $V_{0}$. We present that significant orbital state is more attractive with the increase in $V_{0}$.

\section{Conclusion}
\setcounter{equation}{0}
Based on what is presented in this article, we consider a generalized method for non-hermitic Hamiltonians, so that $PT$ Symmetrical Hamiltonians consist of subsystems named $\eta$-pseudo-hermetic operators if they follow $\eta H=H^{+}\eta$ Similarity conversions. On the other hand in this article, we define two first-order hermitic and non-hermitic operators and it is applied on the Schr\"{o}dinger-like equation (which is obtained from Dirac equation considering constant mass function) to present a general form for $V(x)$ potential which is a function of $F(x)$ super potential which can be used to solve $PT$ symmetrical potentials such as  deformed Woods-Saxon potential. Deformed Woods-Saxon potential is solvable using quantum super symmetry, inherited chain Hamiltonians and shape-invariance traits. Therefore, a new effective potential depending on the diffuseness parameter $"a"$ is used in our computations which eventually helps us in calculating energy eigenvalue.

\pagebreak \vspace{7cm} \pagebreak 
\end{document}